\documentclass[12pt]{article}
\usepackage{amssymb,amsmath,latexsym}
\usepackage{graphicx}  
\usepackage{dcolumn}  
\usepackage{bm}        
\usepackage{amssymb}   
\usepackage{natbib}   
\usepackage{amsmath}
\usepackage{mathtools}
\usepackage{xfrac}
\usepackage{caption}
\usepackage{color}
\usepackage{soul}
\usepackage{hyperref}
\usepackage{float}
\usepackage{lipsum}
\usepackage{enumerate}
\begin{document}
\title{ Why do mayflies change their gill kinematics as they grow? }

\author{R. Chabreyrie$^1$, K. Abdelaziz$^2$, E. Balaras$^1$, K. Kiger$^2$\\
$^1$Department of Mechanical and Aerospace Engineering, \\
The George Washington University, USA\\
\\
$^2$Department of Mechanical Engineering, \\
University of Maryland, USA}

\maketitle
\abstract{The mayfly nymph breathes under water through an oscillating array
  of wing-shaped tracheal gills.  As the nymph grows, the kinematics
  of these gills change abruptly from rowing to flapping. The
  classical fluid dynamics approach to consider the mayfly nymph as a
  pumping device fails in giving clear reasons to this switch.  In
  order to understand the whys and the hows of this switch between the
  two distinct kinematics, we analyze the problem under a Lagrangian
  viewpoint.  We consider that a good Lagrangian transport that
  distributes and spreads water and dissolved oxygen well between and
  around the gills is the main goal of the gill motion. Using this
  Lagrangian approach we are able to provide the reason behind the
  switch from rowing to flapping that the mayfly nymph experiences as
  it grows.  More precisely, recent and powerful tools from this
  Lagrangian approach are applied to in-sillico mayfly nymph
  experiments, where body shape, as well as, gill shapes, structures
  and kinematics are matched to those from in-vivo.  In this letter,
  we show both qualitatively and quantitatively how the change of
  kinematics enables a better attraction, stirring and confinement of
  water charged of dissolved oxygen inside the gills area. From the
  computational velocity field we reveal attracting barriers to
  transport, i.e. attracting Lagrangian coherent structures,
  that form the transport skeleton between and around the gills. In
  addition, we quantify how well the fluid particles and consequently
  dissolved oxgen is spread and stirred inside the gills area.}
%\pacs{47.63.-b, 87.85.gf, 47.61.Ne, 47.52.+j}

Animals have evolved diverse kinematics to generate flows for
locomotion, feeding, cooling and breathing and there seems to be a
trend between these kinematic patterns and the scale of such systems,
as captured by the Reynolds number ($Re=UL/\nu$, where $U$ and $L$
are characteristic velocity and length scales respectivelly, and $\nu$
the kinematic viscosity of the fluid) \cite{Strathmann:1993,
  Walker:2002}.  In particular, the basic trends, which are fairly
well understood at high and low $Re$ indicate that: rowing (i.e. net
flow directed ventrally and mostly parallel to the stroke plane) is
exclusively used at $Re<1$ (see for example \cite{Taylor:1951,
  Gray:1955, Lighthill:1976, Brennen:1977} for flagella and cilia
locomotion), whereas flapping (i.e. net flow directed dorsally and
essentially transverse to the stroke plane) is predominantly used at
$Re>100$ (see \cite{Daniel:1992, Motani:2002, Spedding:2003} for
swimming or flying vertebrates).  At the less studied intermediate
regime $Re=1-20$, \cite{Webb:1986, Daniel:1987, Daniel:1992,
  Fuiman:1997, Walker:2002, Childress:2004}, the same rule of thumb
seems to be valid, when the primary function is the generation of a
propulsive force.  Appendage kinematics also serve other purposes than
locomotion, such as feeding or breathing, where transport (with the
specific goals of attracting, stirring and trapping) maybe more
important than generating a propulsive hydrodynamic force.  Such
systems have received far less attention compared to locomotion, and
there is a wealth of interesting phenomena to explore that may bring
to fruition new, bio-inspired, sensor designs.

\begin{figure}[t]
\begin{center}
\includegraphics[scale=.7]{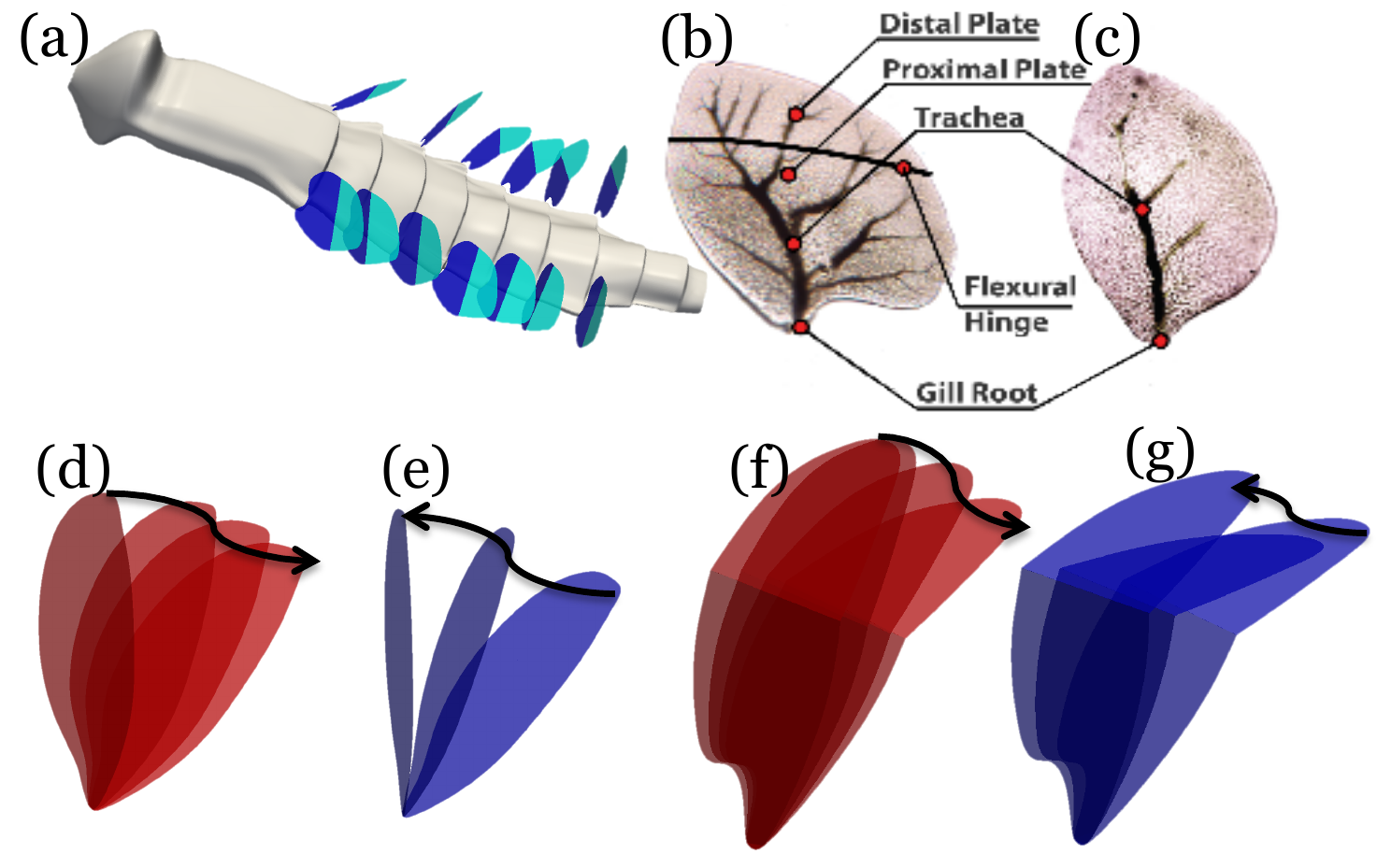}
\end{center}
\caption{(a) In-silico model of mayfly nymph body with its seven pairs
  of gills.  In-vivo gills: (b) mature mayfly; (c) young mayfly.
  In-sillico rowing kinematics for a young mayfly: (d) retraction; (e)
  protraction.  In-sillico flapping kinematics for a mature mayfly:
  (f) retraction; (g) protraction.}
\label{Fig1}
\end{figure}

In the present work we will focus on mayflies, which are insects
belonging to the order of {\textit Ephemeroptera}, referring to their
brief adulthood lifespan of only a few days.  Most of a mayfly's
lifecycle is spent as a nymph in submerged in wetlands, ponds and
river habitats in poorly oxygenated water.  The mayfly nymphs breathe
through tracheal tubes that transport oxygen directly into the tissue
\cite{BRV:BRV181}. In the species considered in this study, these
tracheal tubes branch out to a series of seven pairs of wing-shaped
protrusions known as tracheal gills (see Fig.~\ref{Fig1}). Today it is
established that these gills play a major role in the metabolic
respiration \cite{Babak:1907}.  The water current generated by the
motion of these gills was first observed by \cite{Eastham:1934, Eastham:1936,
  Eastham:1937} in the 30s.  It was not until recently, however, that
more precise quantitative descriptions where reported, which revealed
an interesting feature: as mayfly nymphs grow older they sharply
change their gill kinematics from rowing to flapping
\cite{BIJ:BIJ1314}.  \cite{Sensenig:2010},
conducted detailed quantitative measurements and suggested that the
observed switch is determined by the size of the vortex generated at
the space between the gills.

\cite{Abdelaziz:2013} used a realistic
mayfly model (see Fig.~\ref{Fig1}a) and prescribed the gill kinematics
reported in \cite{Sensenig:2010} (see Fig.~\ref{Fig1}d-g), to conduct
full Navier-Stokes simulations for a range of Reynolds numbers. They
demonstrated that if the mayfly is viewed as a pumping device, then
the resulting mechanical efficiency (i.e. the equivalent to the
mechanical efficiency of a hydraulic pump, which can be derived from
the mechanical energy conservation) is always better for rowing
kinematics at any Reynolds number.  They also examined another pumping
performance parameter, defined as the ratio of the time averaged mass
flow rate towards the mayfly to the time averaged rate of work done by
the gills, which indicated a slight superiority of flapping kinematics
at higher Reynolds numbers.  In this letter, we will consider a
Lagrangian viewpoint, which follows fluid particles as they proceed in
time, to study the breathing mechanism and its relation to the sharp
change in the kinematic patterns.  Our basic hypothesis is that
efficient transport, which attracts and stirs the Lagrangian particles
(and consequently dissolved oxygen) between and around the gills, is
at the origin of the kinematic switch.  In other words, we propose to
view the mayfly as a \textit{stirring} rather than a \textit{pumping}
device as in earlier studies \cite{Sensenig:2010, Abdelaziz:2013}.  An
effective transport/stirring mechanism may be seen as a way to provide
a more efficient way in extracting oxygen from water, which is
particularly relevant in anoxic waters as is the natural environment
for the mayfly.

To study the transport performance we will utilize a dynamical systems
strategy, which during the past two decades has provided new insights
into a variety of fluid mechanics problems \cite{GrigorievBook}, most
notably in small-scale fluid systems for which chaos theory has
provided ways to create efficient and controllable micro-fluidic
mixers \cite{Ottino:90, PhysRevE_77_036314, chabreyrie:072002}.  To
outline the basic principles of the approach let us start from an
Eulerian velocity field, ${\bm V}$, and consider the particle path
equations,
$$ \dot{\bm X}_{{\bm X}_0}= {\bm V}\left({\bm X}_{{\bm X}_0};t\right),$$
where ${\bm X}_{{\bm X}_0}$ represents the location of a Lagrangian
particle at time, $t$, initially located at, ${\bm
  X}_0=\left(x_0,y_0,z_0,t=0\right)$. The solutions of this dynamical
system or the flow $\phi^t$, i.e.~$\phi^t\left({\bm X}_o\right)={\bm
  X}_{{\bm X}_0}(t)$, give all the pathlines (or material lines) of
the fluid flow generated by the velocity field, ${\bm V}$. In our case
${\bm V}$ is obtained from the full Navier-Stokes simulations reported
in \cite{Abdelaziz:2013}, where physiological characteristics
(i.e. body shape, gills shape and kinematics) are carefully matched to
replicate the conditions in the experiments with living mayfly nymphs
reported in \cite{Sensenig:2010}. The in-sillico mayfly used in these
computations is shown in Fig.~\ref{Fig1}a, where one can observe how
the gills are approximated with zero thickness, two-part, hinged
plates.  In the model a perfect bilateral symmetry is assumed for the
mayfly body, gill shapes and motions.

%\section{Cases Studied}
To better understand the physics that lead to the abrupt change in
kinematics we will consider the following cases representing different
stages of the mayflie's life: \textit{i)} an early stage where the
mayfly is around $1-2$ days old; \textit{ii)} a mature mayfly nymph,
where the mayfly is around $40$ days old.  In the former case the
gills resemble small plates with an average length of $0.3$ mm that
can be approximated by a rigid plate as shown in
Fig.~\ref{Fig1}c-d. The Reynolds number is $Re=L^2f\rho/\mu=1.0$, where
$L$, $f$, $\rho$ and $\mu$ are the gill length, beating frequency, fluid density
and viscosity, respectively.  We should note that due to the
lack of gill flexibility, at this $Re$ the reciprocal flapping
kinematics should result in negligible, on average, fluid transport.
In the latter case, the gill size has increased drastically with an
average length of $0.7$ mm and the geometry has become more
complex. In addition a flexural hinge developes as shown in
Fig.~\ref{Fig1}b, f-g, and the gill is now approximated as a two-part
hinged plate.  The Reynolds number in this case is $Re=21.6$.

%%%%%%%%%%%%%%%%%%%%%%%%%%%%%%%%%%%%%%%%%%%%%%%%%%%%%%%%%%%%%%%%%%%%%%%%%%%%%%%%%%%%%%%%%%%%%%%%%%%%
%\section{Results}
First we will focus on identifying dynamically active barriers to
transport, labeled as Lagrangian Coherent Structures (LCS)
\cite{haller:99, Haller2000352, G2001248, haller:3365}.  These
structures are now seen to be crucial in understanding transport
phenomena in a variety of time-dependent systems ranging from large
scale, e.g. oceanic \cite{Lekien:2005, Irina:2010} or atmospheric
\cite{tang:017502} flows, to small scale, e.g. hemodynamics
\cite{shadden:017512, arzani:081901, Shawn:2012a, CNM:CNM2523} or
swimming \cite{348461379,0953-8984-21-20-204105}.  These structures
divide the fluid into dynamically distinct regions, revealing features
hidden in the velocity field.  In other words, these LCS act as
attractive or repulsive barriers to transport for fluid particles.
These attractive or repulsive LCS can be defined as the ridges of the
Finite Time Lyapunov Exponent (FTLE) map calculated backward or
forward in time. This map consists of associating an FTLE,
$\mathcal{L}$, with an initial condition, ${\bm X}_{0}$.  The Lyapunov
exponent can be seen as a measure of how two trajectories, starting
initially close from each other, diverge.  First, in order to compute
the FTLE map, we consider the tangent flow
$$ \dot{{\bm J}}^t={\bm\nabla}{\bm V}{\bm J}^t, $$
where ${\bm J}^t$ and ${\bm \nabla}{\bm V}$ are the Jacobian and
matrix of variations, respectively.  The initial condition is ${\bm
  J}^0={\bm I}$, where ${\bm I}$ is the three-dimensional identity
matrix.  $J^t(\bm{X}_0)={\partial \phi^t\left({\bm X}_0\right)}
/{\partial {\bm X_0}}$ describes the deformation at time $t$ of an
infinitesimal sphere of neighboring initial conditions starting at
$\bm{X}_0$.  Then, the FTLE map for a time, $t=\tau$, is computed as
$$\mathcal{L}({\bm X}_{0},\tau)=\frac{\ln|\gamma_{max}({\bm X}_{0})|}{|\tau|},$$
where $\gamma_{max}$ is the largest eigenvalue (in norm) of
$J^{\tau}$.  Finally, in order to reveal the LCS, ridges from the FTLE
map, ${\bm X}_0\rightarrow \mathcal{L}({\bm X}_0,\tau)$, are extracted
for a given time, $\tau$.  It is important to note that the LCS
indicate a strong attractive or repulsive transport (i.e. local
maximum in Lyapunov exponent map) on a thin restrained area
(i.e. ridges). As the time of integration, $\tau$, is increased this
fine line or surface extends throughout the fluid space. Consequently,
these LCS obtained by using the ridges of the FTLE map are localized
where the most effective attractive and repulsive dynamics occur in
the all fluid space.  Since, in this work mayfly naids are seen as
short time particle capturing/attracting systems, we will focus on
capturing attractive LCS after a few oscillating periods of the gills.

\begin{figure}
\begin{center}
\includegraphics[scale=0.8]{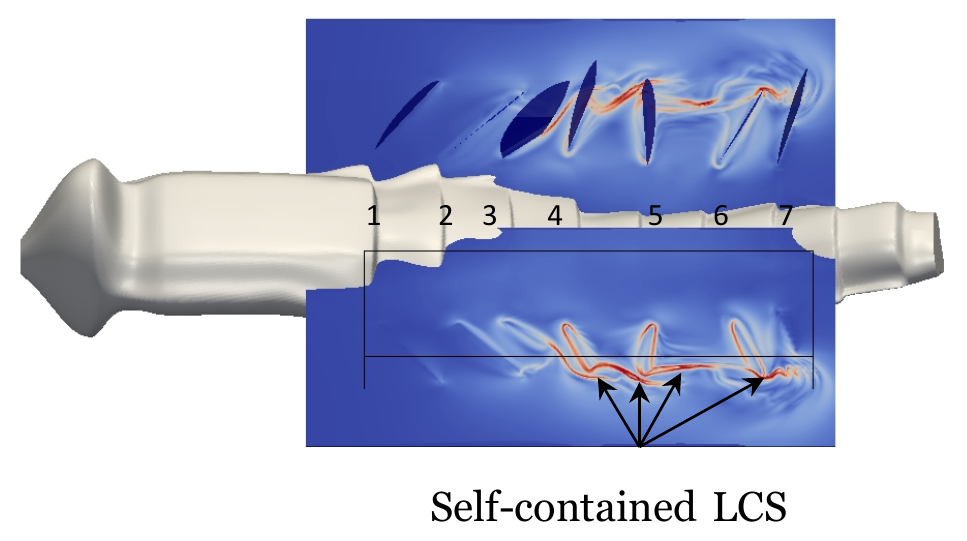}
\end{center}
\caption{FTLE field computed backward in time over five periods of
  gill oscillations for a young nymph ($Re=1.0$) with rowing
  kinematics. For clarity we only plot two sections from the
  three-dimensional volume data: a horizontal section (bottom part),
  and an oblique section (top part). The attracting LCS is revealed by
  the ridges of the FTLE. Gills are shown in solid blue color.}
\label{Fig2a}

\end{figure}
\begin{figure*}
\begin{center}
\includegraphics[scale=0.8]{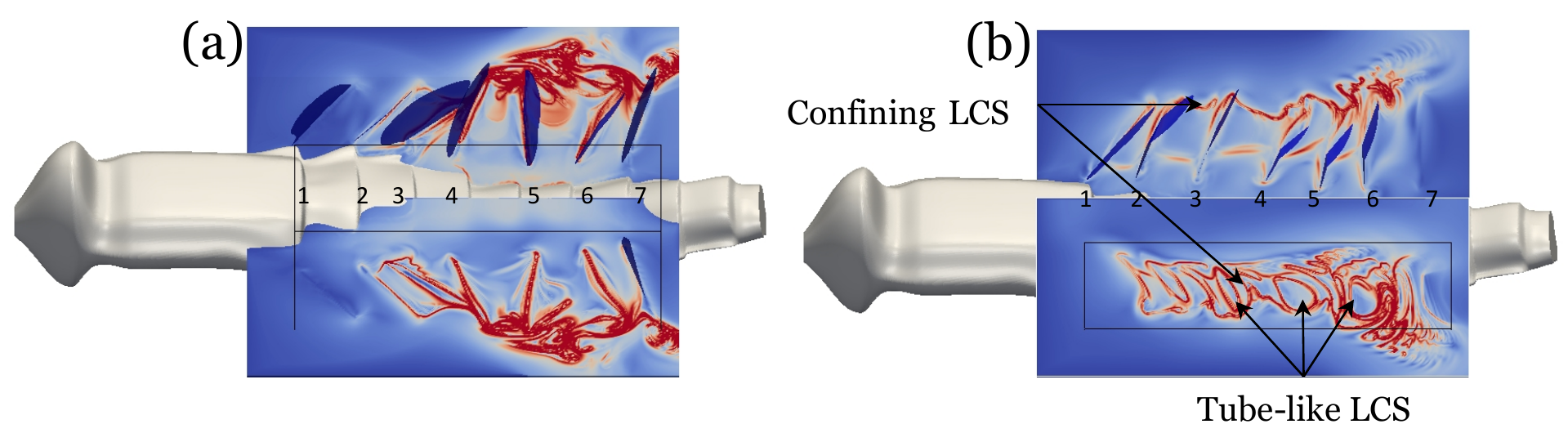}
\end{center}
\caption{FTLE field computed backward in time over five periods of
  gill oscillations for mature nymph ($Re=21.6$) with: (a) rowing
  kinematics; (b) flapping kinematics. For more details see caption of
  Fig.~\ref{Fig2a}.}
\label{Fig2bc}
\end{figure*}

Figures~\ref{Fig2a} and \ref{Fig2bc} display the FTLE field for:
\textit{i)} a young nymph at $Re=1.0$ and, \textit{ii)} a mature nymph
at $Re=21.6$.  For both cases, the seven pairs of gills generate a
complex three-dimensional unsteady velocity field, which in turn
produces a very complex three-dimensional transport with embedded
features.  In order to clearly reveal these key features we plot two
sections from the three-dimensional field: a horizontal and an oblique
section shown at the bottom and top parts of Figs~\ref{Fig2a} and
\ref{Fig2bc} respectively. For each kinematic pattern these two
sections are specificaly selected to illustrate where the structures
are the most relevant for transport.

\paragraph{i)} In the case of a young nymph ($Re=1.0$) with the rowing
kinematic (the only gill motion possible), we see sharp ridges in the
FTLE field corresponding to strong attractive LCS.  From the
horizontal and oblique sections of Fig.~\ref{Fig2a}, we see that these
LCS are mainly localized between gill pairs $3-6$.  Each gill
generates its own self-contained attractive LCS going only along each
respective gill (see horizontal section in Fig.~\ref{Fig2a}) and then
away from its tip with little transport generated in the inter-gill
space (see oblique section in Fig.~\ref{Fig2a}).  Such a transport
structure is a direct consequence of the fact that with rowing
kinematics, each gill behaves as an independent (i.e. without
interaction with its neighboring gills) single plate that mainly
pushes fluid along and away from itself.

\paragraph{ii)} For a mature nymph ($Re=21.6$), both the rowing and
flapping kinematics are possible.  In the rowing case, although the
LCS have been developed due to the increase of the Reynolds number,
the transport structure is qualitatively the same as in the young
nymph case.  Each of the central gills $3-6$ generates its own
independent LCS that goes along each respective gill (see horizontal
section in Fig.~\ref{Fig2bc}a) and escape the gill area slightly above
the mayfly body (see oblique section in Fig.~\ref{Fig2bc}a).  The main
difference with the young nymph is an extension of LCS from their gill
tips to outside of the gill area (solid black lines in
Fig.~\ref{Fig2bc}a).  Such an extension of the LCS does not provide
better attraction or capture for the gills. It simply provides a
faster ejection of Lagrangian particles from the intra-gill area.  In
the flapping case, the transport structures change radically from the
rowing cases.  Fig.~\ref{Fig2bc}b clearly shows very neat ridges in
the FTLE field, i.e. very strong LCS that display an intricate
braiding of the transport structures, revealing two main features:
From the top oblique section in Fig.~\ref{Fig2bc}b, one can observe
that the gills generate a {\it confining LCS} structure. This
confining LCS starts from gill pair $2$ and encapsulates the gill area
by joining the other gill tips and filling the inter-gill space.  The
other interesting feature is revealed from the horizontal section in
Fig.~\ref{Fig2bc}b, where one can observe \textit{tube-like LCS}
between gill pairs $3-4$, $4-5$, $5-6$ (see annotations in
Fig.~\ref{Fig2bc}a).  Such tube-like structures can be seen as {\it
  dynamical pockets} where Lagrangian particles are trapped.  These
two features indicate a transport mechanism of confinement of fluid
particles, between and around the gills, which can be seen as a way to
capture and extract more efficiently the dissolved oxygen in water.

From these qualitative observations, we can already understand the
advantages of switching from rowing to flapping.  When flapping
kinematics is possible, i.e. gill flexibility and inertial effects are
strong enough, elaborate transport structures are produced, which
attract, stir and trap fluid particles between and around the gills.
Although the qualitative results above point to the primary reason in
the change of kinematics as the nymph grows, it is also important to
provide quantitative measures of how well Lagrangian particles are
stirred between and around the gills.  Here we will quantify the
degree of stirring as a function of spatial location by introducing
the stirring index, $M$, through the box counting
method~\cite{Stremler:08}. This technique offers the advantage of
being relatively easy to implement, and computationaly inexpensive.
Let us follow $N_p$ Lagrangian particles, and divide the domain
enclosing the gills into, $N_x \times N_y \times N_z$ boxes (see black
edges in Figs.~\ref{Fig2a},~\ref{Fig2bc}).  At each time, $t$, the
number of particles, $n_i$, inside each box, $i$, is computed and then
the particle rate, $r_i$, is calculated as follows:
$$
r_i=\frac{n_i}{n_p}~\mbox{ if } n_i<n_p, ~~r_i=1~\mbox{ if } n_i\geq
n_p,
$$
where $n_p$ is the average number of Lagrangian particles,
i.e.~$n_p=N_p/(N_x N_y N_z)$.  After computing the fraction of
particles in each box and at each time, the time evolution of the
stirring index, $M(t)$, is calculated by taking the average of $r(t)$
over all boxes:
$$
M(t)=\frac{1}{N_x N_y N_z}\sum_{i=1}^{N_x N_y
  N_z}r_i(t),~~\mbox{with}~M(t)\in]0,1[.
$$
The case with the highest stirring index value corresponds to the case
for which Lagrangian particles are spread widely and in the most
uniform way inside the gill area.  Such conditions enhance the
spreading of dissolved oxygen in the water inside the gill area, and
therefore enhance breathing performance.
\begin{figure}[ht]
\begin{center}
\includegraphics[scale=.6]{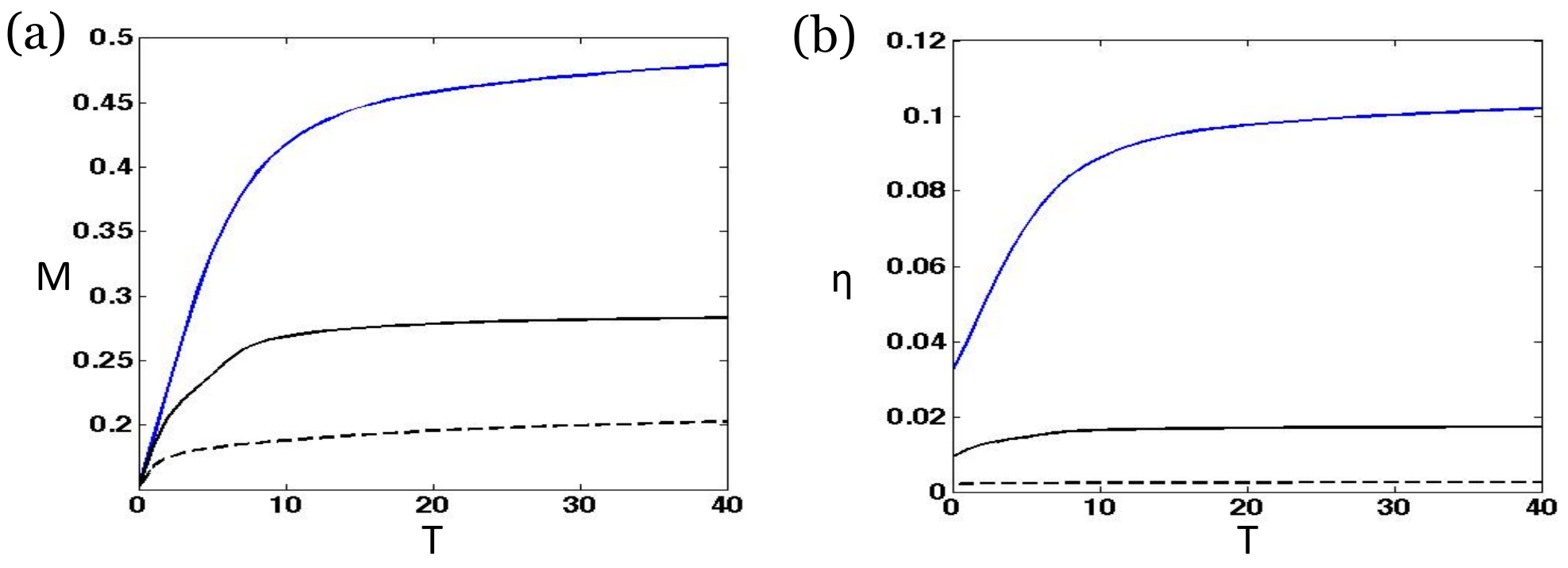}
\end{center}
\caption{\label{Fig3} a) Stirring index and b) performance versus time
  computed over forty gill oscillation cycles, under $N_{x} \times
  N_{y} \times N_{z}=45 \times 50 \times 150$, for $N_p=10^8$
  particles. Young nymph rowing kinematics (dashed
  line);  mature nymph with rowing (solid black line ) and flapping
  (solid blue line) kinematics.}
\end{figure}
The stirring index, $M$, is shown in Fig.~\ref{Fig3}a for the three
cases considered, i.e.  young nymph ($Re=1.0$) with rowing kinematics,
mature nymph ($Re=21.6$) with both rowing and flapping kinematics.
For the young nymph with rowing kinematics after a small increase, $M$
ends up on a very small plateau around $0.2$, indicating poor
stirring.  For the mature nymph with rowing kinematics we see that
after a moderately fast increase phase, $M$ reaches a pseudo plateau
at medium height $0.28$.  For mature nymph with flapping kinematics,
on the other hand, $M$ shows a drastic increase in the stirring level
all over the $40$ periods considered.  In particular, very steep
growth is observed in less than $5$ periods, where the value of $M$
reaches a high plateau of around $0.45$. This is almost two times
higher than the plateau attained by the rowing kinematics.  This  result
reinforces the qualitative observations made earlier with the
stransport structures: the switch to flapping kinematics enable the
production of a transport mechanism that traps and stirs more than the
crude and simple pumping transport generated by the rowing of the
gills.  The advantage of kinematics switching can even be more
revealed by looking at the performance of the stirring.  The stirring
performance parameter, $\eta$, can be defined as the ratio between stirring index,
$M$, and the average rate of work done by the gills over one period of
oscillation:
$$
\eta=\sfrac{M(t)}{~\overline{\dot{W}}_{gills}},
$$
where the rate of work $\dot{W}_{gills}$, is computed as:
\begin{equation}
 \dot{W}_{gills}= \int \limits_{gills}\bm{V} \bm{\tau} d\bm{A}, \;\;\;
\bm{\tau}=-P\bm{I}+\mu(\nabla\bm{V}+\nabla\bm{V}^T).
\label{eq:work}
\end{equation}
In the definition of the stress tensor, $\tau$, in Eq.~(\ref{eq:work})
above, $P$, $\bm{I}$ and $\bm{V}$ are the pressure, identity matrix and the velocity vector, respectively.
 $\overline{\dot{W}}_{gills}$ has been nondimensionalized by the gill length $L$, beating frequency $f$, fluid desnity $\rho$ and viscosity $\mu$.
Fig.~\ref{Fig3}b displays the values of $\eta$ over forty periods of
gill oscillation for the three cases considered.  For the young nymph
$\eta$ is extremely low $<.002$ showing very inefficient stirring
transport.  For the mature nymph with rowing kinematics, $\eta$ is
slightly higher but still represents a highly inefficient stirring
system.  As for the mature nymph with flapping kinematics $\eta$,
increases excessively and reaches a plateau almost an order higher
than in the rowing case at around $\eta \sim 0.1$.
%%%%%%%%%%%%%%%%%%%%%%%%%%%%%%%%%%%%%%%%%%%%%%%%%%%%%%%%%%%%%%%%%%%%%

%\section{Conclusion}
In summary, this letter provides evidence on what causes the switch in
gill kinematics as mayfly nymphs grow.  Using a Lagrangian viewpoint, 
and looking at the mayfly nymph as a stirrer, we were able to
show both qualitatively and quantitatively the advantages in switching
from rowing to flapping as they grow.  In particular, through the
attracting LCS, we have qualitatively revealed that flapping
kinematics generate a complex transport structure that attracts, stirs
and confines around the gills, providing better access to the
dissolved oxygen present in water.  We reach the same conclusion by
quantifying how well the fluid particles are stirred between the gills
via the box counting method.

Overall for young nymphs, due to the lack of gill flexibility and
inertial effects, rowing is the only way to create non-negligible
fluid transport and generate a water current towards their
body/gills. Consequently, through this simplistic transport structure,
the gills can absorb some of the dissolved oxygen present in the
water.  In a mature nymph, on the other hand, the gills develop
flexion lines, and due to their size inertial effects become
important. In this case flapping kinematics generate a complex
transport structure that greatly enhances oxygen extraction.
 
%%%%%%%%%%%%%%%%%%%%%%%%%%%%%%%%%%%%%%%%%%%%%%%%%%%%%%%%%%%%%%%%%%%%%

%\section*{Acknowledgments}
Support from the National Science Foundation, CBET-1067066, is
gratefully acknowledged.

\bibliographystyle{apsrmp4-1}
\bibliography{biblioIII}
\end{document}